# On Identification of Sparse Multivariable ARX Model: A Sparse Bayesian Learning Approach [★]

J. Jin, Y. Yuan[†], W. Pan, D. L. T. Pham, C. J. Tomlin, A. Webb, J. Goncalves

**Abstract**

Although system identification of complex networks is widely studied, most of the work focuses on modelling the dynamics of the ground truth without exploring the topology because in many applications, the network topology is known as a priori. For instance, in industry since the system is artificial, its topology is fixed and the system dynamics is identified for the purpose of control. Nevertheless, in other cases, especially natural sciences, finding the correct network topology sometimes is even more important than modelling potential dynamics since it helps scientists understand the underlying mechanism of biological systems. More importantly, one of the most crucial features of such networks is sparsity which causes a huge difference if no effort is put on exploring the sparse structure. To fill the gap, this paper begins with considering the identification of sparse linear time-invariant networks described by multivariable ARX models. Such models possess relatively simple structure thus used as a benchmark to promote further research. With identifiability of the network guaranteed, this paper presents an identification method that infers both the Boolean structure of the network and the internal dynamics between nodes. Identification is performed directly from data without any prior knowledge of the system, including its order. The proposed method solves the identification problem using Maximum a posteriori estimation (MAP) but with inseparable penalties for complexity, both in terms of element (order of nonzero connections) and group sparsity (network topology). Such an approach is widely applied in Compressive Sensing (CS) and known as Sparse Bayesian Learning (SBL). We then propose a novel scheme that combines sparse Bayesian and group sparse Bayesian to efficiently solve the problem. The resulted algorithm has a similar form of the standard Sparse Group Lasso (SGL) while with known noise variance, it simplifies to exact re-weighted SGL. The method and the developed toolbox can be applied to infer networks from a wide range of fields, including systems biology applications such as signaling and genetic regulatory networks.

*Key words:* System Identification; Sparse Bayesian Learning; Polynomial Model.

## 1 Introduction

In system identification, the prediction error method (PEM) equivalent to Maximum-likelihood (ML) is well developed. It is applicable to a big family of black-box models such as ARX, ARMAX and Box-Jenkins. Given an infinite number of data points and some mild assumptions, PEM attains consistent and asymptotically estimates [11]. However, the main disadvantage for PEM is that for most model structures it has to deal with non-linear numerical optimization for the parameter estimation, which normally only guarantees local convergence. Some exceptions exist for the models in linear regression form such as ARX and FIR model whose optimization problem is solved analytically [28]. The second issue is that without any prior knowledge about the system order, over-fitting is essentially inevitable. Hence, most proper model is selected afterwards using AIC or BIC criterion. As such, the procedure demands more computation effort than necessary. Finally, when dealing with MIMO system, assuming no prior knowledge of the topology, PEM generates full transfer matrices even if the ground truth is sparse. The advanced technique, MAP (Type I method), thus is proposed to penalize the parameters. While it effectively relieves over-fitting, the weighting variable which controls the trade-off between data fitting and model complexity is evaluated by cross-validation which increases the computation burden and causes information waste.

In the machine learning literature, a well developed technique called Sparse Bayesian Learning (Type II method) actually is a general form of MAP and suitable to solve linear regression problems [1,5,20,25]. Since it applies in-

[★] Junyang Jin and Alex Webb are with Circadian Signal Transduction Group, Department of Plant Sciences, University of Cambridge. Ye Yuan is with School of Automation, Huazhong University of Science and Technology. Wei Pan is with Cardwell Investment Tech and Imperial College London. Duong L.T. Pham is with the Luxembourg Centre for Systems Biomedicine. Claire J. Tomlin are with the Department of Electrical Engineering and Computer Sciences, UC Berkeley. Jorge Gonçalves is with the Department of Engineering, University of Cambridge and the Luxembourg Centre for Systems Biomedicine. [†]For correspondence, yye@hust.edu.cn.

separable prior, Type I method is a special case of SBL. Compared with conventional Type I method, SBL has superior performance and is tuning-free [25]. In previous research, it is already applied to solve nonlinear system identification [14, 15]. In terms of the model structure of nonlinear systems, element SBL or group SBL (GSBL) is utilized alone to cope with different data set. For a general SISO system, approximation of an IIR system by a FIR one using truncation usually is necessary and the kernel method which generalizes SBL is proposed [6]. To infer a MIMO system described by a dynamical structure function (DSF), only the topology of the network is inferred by Lasso algorithm using steady state time-series data while internal dynamics remains unknown [10]. A similar work is done to infer the multivariable ARX model with known polynomial order using Block Orthogonal Matching Pursuit (BOMP) which focuses more on identifying network topology [16]. A non-parametric technique via Bayesian approach is developed to infer the Granger causality and intended to get correct topology and good prediction [7].

In this paper, we consider the identification of a network described by a multivariable ARX model which has the simplest model structure in the MIMO family. We also assume no prior knowledge of the system order and network topology. By the virtue of the model structure, SBL is introduced to solve this identification problem. To promote sparse network topology and search for lowest order possible, whereas Sparse Group Lasso (SGL) can be used to solve the same problem as a Type I method, we propose a novel scheme to combine SBL and GSBL. The resultant algorithm indicates such approach is closely linked to SGL method.

The paper is organized as follows. Section 2 introduces the polynomial model and discusses its identifiability. Section 3 formulates the network reconstruction problem. Section 4 solves the problem of network identification using a combination of SBL and GSBL. Section 5 considers an extended nonlinear polynomial model. Section 6 applies the method to a randomly generated network and shows how its topology and internal dynamics can be reconstructed accurately. Finally, Section 7 concludes and discusses further development in this field.

*Notation*: The notation in this paper is standard. $I$ denotes the identity matrix. If $L \in R^{n \times n}$, $diag\{L\}$ denotes a vector which consists of diagonal elements of matrix $L$. If $l \in R^n$, $diag\{l\}$ denotes a diagonal matrix with its diagonal elements to be the vector $l$. $trace\{L\}$ denotes the trace of matrix $L$. A matrix $L \succeq 0$ means $L$ is positive semi-definite. A vector $v \geq 0$ means each element of the vector is non-negative. A vector $v^k$ means array power of $k$ of $v$. A vector $y(t_1 : t_2)$ denotes a row vector $\begin{bmatrix} y(t_1) & y(t_1+1) & \cdots & y(t_2) \end{bmatrix}$. $(.^*)$ means array multiplication.

## 2 MODEL FORMULATION

The network considered is described by a multivariable ARX model: $A(z^{-1})Y(t) = B(z^{-1})U(t) + E(t)$, where

$$A(z^{-1}) = I + A_1 z^{-1} + ... + A_{n_a} z^{-n_a}, \\ B(z^{-1}) = B_1 z^{-1} + ... + B_{n_b} z^{-n_b}. \quad (1)$$

$Y(t) \in R^p$ are the nodes of the network, $U(t) \in R^m$ denotes input, and $E(t) \in R^p$ is i.i.d white Gaussian noise. $A(z^{-1})$ is a polynomial matrix showing the connectivity of each node to other nodes including self-loops. Similarly, $B(z^{-1})$ is a polynomial matrix relating the input to the nodes. The boolean structure of the network is reflected by the nonzero elements in $A(z^{-1})$ and $B(z^{-1})$, whereas the dynamics of the network are given by the elements in these transfer matrices. The multivariable ARX model defines the network topology and results in a unique Input-Output map:

$$Y(t) = G(z^{-1})U(t) + H(z^{-1})E(t), \quad (2)$$

where

$$G(z^{-1}) = A^{-1}(z^{-1})B(z^{-1}) \\ H(z^{-1}) = A^{-1}(z^{-1}). \quad (3)$$

Another useful way to describe a linear network is by dynamical structure function (DSF) which is uniquely defined by ARX model in this case.

$$Y(t) = Q(z^{-1})Y(t) + P(z^{-1})U(t) + H(z^{-1})E(t), \quad (4)$$

where

$$Q(z^{-1}) = I - \tilde{A}^{-1}(z^{-1})A(z^{-1}) \\ P(z^{-1}) = \tilde{A}^{-1}(z^{-1})B(z^{-1}) \\ H(z^{-1}) = \tilde{A}^{-1}(z^{-1}) \\ \tilde{A}(z^{-1}) = diag\{diag\{A(z^{-1})\}\}. \quad (5)$$

**Remark 1** : *In general, the map between DSF and Input-Output transfer matrix is not bijective. A DSF uniquely defines a transfer matrix whereas a transfer matrix corresponds to infinite many DSF. Given a transfer matrix, its associated DSF is unique if and only if partial information is available about the matrix $\begin{bmatrix} P(z^{-1}) & H(z^{-1}) \end{bmatrix}$ [9, 22, 26]. For an ARX model, since each node is only influenced by its own intrinsic noise, the transfer matrix $H(z^{-1})$ is diagonal so that DSF is unique.*

The theory of identifiability of a general black-box SISO system is clear [11]. In addition to some necessary conditions to the model structure, not all but at least one polynomial order of rational transfer functions must coincide with the ground truth. The identifiability of a general MIMO system is more subtle. It is discussed with specific model classes instead of a general model structure.

Also, the conditions vary a lot under different parameterization schemes. Nevertheless, we find the identifiability of a multivariable ARX model is straightforward.

**Lemma 2** : *Consider the multivariable ARX model in eq. (2). The degrees of polynomials are $n_a \geq n_a^*$ and $n_b \geq n_b^*$ where $\theta^*$, $n_a^*$ and $n_b^*$ are ground truth. Then $\theta^*$ corresponds to a globally identifiable $\theta$ of the model structure.*

The proof is straightforward. Given $H(z^{-1})$, $A(z^{-1})$ is fixed which in turn deduces $B(z^{-1})$ given $G(z^{-1})$. The lemma indicates an important point that no prior knowledge of the polynomial order is needed aside from an upper bound to guarantee the identifiability of the model structure.

## 3 RECONSTRUCTION PROBLEM FORMULATION

We parameterize each node of multivariable ARX model structure in the same form and set a corresponding upper bound to the polynomial orders of ground truth as:

$$y_i(t) = A_{i1}(z^{-1})y_1(t) + \ldots + [1 - A_{ii}(z^{-1})]y_i(t) \\ + \ldots + B_{im}(z^{-1})u_m(t) + e_i(t). \quad (6)$$

$y_i(t)$ denotes $i$th node, $u(t)$ input and $e_i(t)$ i.i.d Gaussian noise and:

$$\begin{aligned} A_{ii}(z^{-1}) &= a_1^{ii} z^{-k} + a_2^{ii} z^{-k+1} + \ldots + a_k^{ii} z^{-1} + 1 \\ A_{ij}(z^{-1}) &= a_1^{ij} z^{-k} + \ldots + a_{(k-1)}^{ij} z^{-2} + a_k^{ij} z^{-1} \\ B_{ij}(z^{-1}) &= b_1^{ij} z^{-k} + \ldots + b_{(k-1)}^{ij} z^{-2} + b_k^{ij} z^{-1} \\ k &\geq max\{n_a^*, n_b^*\}, \end{aligned} \quad (7)$$

where $a$ and $b$ denote the parameters in polynomial matrices $A(z^{-1})$ and $B(z^{-1})$ respectively, superscript $ij$ the polynomial of $ij$th element of matrices and subscript $i$ the index of $i$th coefficient of polynomial.

Assume that time-series data collected from discrete time indices 1 to $t$ for each node and input are available. For the $i$th node, we define following matrices and vectors:

$$y = \begin{bmatrix} y_i(t) \\ \vdots \\ y_i(k+1) \end{bmatrix}, w = \begin{bmatrix} w_1 \\ \hline \vdots \\ \hline w_{p+m} \end{bmatrix}$$

$$\Phi = \begin{bmatrix} y_1(t-k:t-1) & \ldots & -y_i(t-k:t-1) \\ \vdots & \ldots & \vdots \\ y_1(1:k) & \ldots & -y_i(1:k) \end{bmatrix} \quad (8)$$

$$w_j = \begin{cases} \begin{bmatrix} a_1^{ij} & \ldots & a_k^{ij} \end{bmatrix}^T & \text{if } j \leq p-1, j \neq i \\ \begin{bmatrix} b_1^{ij} & \ldots & b_k^{ij} \end{bmatrix}^T & \text{if } p \leq j \leq p+m-1 \\ \begin{bmatrix} a_1^{ii} & \ldots & a_k^{ii} \end{bmatrix}^T & \text{if } j = p+m \end{cases}$$

$$\lambda = E\{e_i(t)^2\}.$$

The likelihood function based on Baye's rule is thus:

$$\begin{aligned} p(y|w, \lambda) &= \prod_{l=0}^{t-k-1} p(y_i(t-l) | Y_{t-l-1}^{t-l-k}, w, \lambda) \\ &= \frac{1}{(2\pi\lambda)^{(t-k)/2}} \exp\{\frac{1}{2\lambda}\|y - \Phi w\|_2^2\} \end{aligned} \quad (9)$$

where $Y_{t2}^{t1}$ means data points of all the nodes and input from $t_1$ to $t_2$.

**Remark 3** : *The likelihood function of multiple independent experiment is just the multiplication of likelihood of each experiment. If the experiment condition does not change too much so that noise variance is same, the likelihood function is in the same form of (9) whose $y$ and $\Phi$ are stack of counterparts of individual experiment. Even if this assumption does not apply, only little modification is needed to fit the framework in this paper.*

Note that this likelihood distribution is not presented in a standard form of Gaussian distribution, which however shows its logarithm is a quadratic function of $w$. By maximizing its logarithm with respect to $w$, we end with the PEM (ML) method. With infinite data points, PEM is guaranteed to converge to the ground truth model [11]. In practice, given limited data, PEM may suffer from over-fitting due to the upper bound of polynomial orders and also generate a full connected network even if the true one is sparse. Therefore, penalties for both network topology and model complexity are essential. Referring to the parameterizations in eq. (8), sparse network can be interpreted as group sparse $w$ whereas sparsity within each group indicates reduced order of polynomials. This consideration naturally leads us to the usage of Maximum a-posteriori method (MAP). A direct framework to achieve those two levels of sparsity is Sparse Group Lasso (SGL) as a Type I MAP method. In this paper, we resort to Sparse Bayesian Learning as a Type II MAP method. It employs inseparable prior distribution which generalizes the penalty of Type I. It is also shown to be superior than most Type I MAP schemes. In addition, we show later that our approach is closely related to SGL.

## 4 RECONSTRUCTION VIA SPARSE BAYESIAN LEARNING

### 4.1 Sparsity inducing priors

Full Bayesian treatment requires to introduce a prior distribution for $w$. We define a distribution $p(w)$ in a general form as: $p(w) \propto \exp\left[-\frac{1}{2}\sum_j g(w_j)\right]$. If $w$ has certain sparse properties, a prior inducing sparsity like Generalized Gaussian, Student's t and Logistic is assigned to $p(w)$. The function $g(\cdot)$ of such priors is usually a concave, non-decreasing function of $|w_j|$ [25]. Estimating $w$ using its posteriori mean is intractable in this case because the posterior distribution $p(w|y)$ is non-Gaussian and not analytical. To simplify the problem, sparse Bayesian learning approximates $p(w|y)$ with a Gaussian distribution so that the solution, $E(w|y)$ can be easily calculated. To do that, first note that virtually all sparse priors namely super Gaussian can be presented in a variational form which yields a lower bound for the sparse prior $p(w)$. There are two types of variational representations; while we apply the convex type which is more general than the integral one [12, 21].

To induce element sparsity to $w$, we introduce a prior $p(w)$ as [14, 25]:

$$\begin{aligned} p(w) &= \prod_{i=1}^{p+m} p(w_i) = \max_{\beta \geq 0} \mathcal{N}(w|0, B)\varphi(\beta) \\ p(w_i) &= \prod_{j=1}^{k} p(w_{ij}) = \max_{\beta_i \geq 0} \mathcal{N}(w_i|0, B_i)\varphi(\beta_i), \\ p(w_{ij}) &= \max_{\beta_{ij} \geq 0} \mathcal{N}(w_{ij}|0, \beta_{ij})\varphi(\beta_{ij}), \end{aligned} \quad (10)$$

where subscript $i$ denotes $i$th group in a vector and $j$ the $j$th element in that group. $\beta$ is a vector of hyper-parameters which controls element sparsity of the vector $w$. $B$ is the covariance matrix of Gaussian distribution and parameterized by vector $\beta$ as $B = \text{diag}\{\beta\}$.

For group sparsity, the corresponding prior is [15, 27]:

$$\begin{aligned} p(w) &= \prod_{i=1}^{p+m} p(w_i) = \max_{\gamma \geq 0} \mathcal{N}(w|0, \Gamma)\varphi(\gamma) \\ p(w_i) &= \max_{\gamma_i \geq 0} \mathcal{N}(w_i|0, \gamma_i I)\varphi(\gamma_i). \end{aligned} \quad (11)$$

where $\gamma$ is a vector of hyper-parameters which controls group sparsity of $w$.

**Remark 4** : *If the true model is ARX type, then the last group of $w$ which presents auto-regression can be excluded from the group sparsity to improve the estimation accuracy. If not, the resultant model can be an FIR OE model.*

To promote both element and group sparsity, we multiply (10) by (11) and normalize it to get a proper distribution:

$$p(w) = C * \max_{\gamma \geq 0, \beta \geq 0} \mathcal{N}(w|0, B) \mathcal{N}(w|0, \Gamma)\varphi(\beta)\varphi(\gamma) \quad (12)$$

where $C$ is the normalization constant and can be absorbed by functions $\varphi(\beta)$ or $\varphi(\gamma)$. Hence, we get a lower bound of this prior as:

$$\begin{aligned} p(w) &\geq \mathcal{N}(w|0, B)\mathcal{N}(w|0, \Gamma)\varphi(\beta)\varphi(\gamma) \\ &= \hat{p}(w). \end{aligned} \quad (13)$$

If either $\beta_{ij}$ or $\gamma_i$ approaches 0, the corresponding Gaussian distribution becomes a Dirac delta function imposing element or group sparsity to $w$.

**Remark 5** : *The conventional way to promote both element and group sparsity is to use hierarchical Bayesian by introducing two hyperparameters where one is regulated by the other. The main problem is how to choose corresponding hyper-distributions, which is non-trivial. Also, the hyperparameter which is deeper in hierarchy has less impact on the inference procedure [8]. That means the resultant penalty is probably not strong enough to impose group sparsity to the weighting vector $w$. As such, multiplying two priors makes sense since both hyperparameters influence $w$ directly. Nevertheless, it is shown later that this scheme can be transferred into an equivalent hierarchical Bayesian format.*

### 4.2 Type II Maximization

Although the implicit prior $\hat{p}(w)$ is improper, we can still get a normalized posteriori distribution of $w$ as:

$$\hat{p}(w|y) = \frac{p(y|w)\hat{p}(w)}{\int p(y|w)\hat{p}(w)dw}. \quad (14)$$

Clearly, $\hat{p}(w|y)$ is Gaussian since $log(p(w|y))$ is a quadratic function of $w$:

$$\hat{p}(w|y) = \mathcal{N}(w|\mu, \Sigma), \quad (15)$$

where

$$\begin{aligned} \Sigma &= \left[(\Gamma^{-1} + B^{-1}) + \lambda^{-1}\Phi^T\Phi\right]^{-1} \\ \mu &= \lambda^{-1}\left[(\Gamma^{-1} + B^{-1}) + \lambda^{-1}\Phi^T\Phi\right]^{-1}\Phi^T y. \end{aligned} \quad (16)$$

Now, the core of the problem is to choose proper hyperparameters $\beta$ and $\gamma$ so that $\hat{p}(w|y)$ is close to the real

$p(w|y)$ under the selected criterion. One way is to minimize the misaligned mass between $p(w)$ and $\hat{p}(w)$ weighted by the marginal likelihood $p(y|w)$ which is also called evidence maximization or Type II maximization [2, 14, 25]. It is equivalent to estimating hyperparameters using the maximum likelihood method:

$$\begin{aligned}(\gamma^*, \beta^*, \lambda^*) &= \arg \min_{\beta,\gamma,\lambda \geq 0} \int p(y|w)|p(w) - \hat{p}(w)|dw \\ &= \arg \min_{\beta,\gamma,\lambda \geq 0} -2\log \int p(y|w)\hat{p}(w)dw \\ &= \arg \min_{\beta,\gamma,\lambda \geq 0} -2\log \ \hat{p}(y|\beta,\gamma,\lambda)\end{aligned} \quad (17)$$

**Proposition 6** : Following (17), $w$, hyperparameters and noise covariance, $\lambda$ can be estimated by solving the optimization problem:

$\mathcal{L}_1$ :
$$\min_{\beta,\gamma,\lambda,w} \lambda^{-1}\|y - \Phi w\|_2^2 + w^T(\Gamma^{-1} + B^{-1})w$$
$$+ \log |B + \Gamma| + log|\lambda I + \Phi(\Gamma^{-1} + B^{-1})^{-1}\Phi^T|$$
Subject to:
$$\beta \geq 0, \gamma \geq 0, \lambda \geq 0$$
Or: (18)
$\mathcal{L}_2$ :
$$\min_{\beta,\gamma,\lambda} y^T \left[\lambda I + \Phi(\Gamma^{-1} + B^{-1})^{-1}\Phi^T\right]^{-1} y$$
$$+ \log |B + \Gamma| + log|\lambda I + \Phi(\Gamma^{-1} + B^{-1})^{-1}\Phi^T|$$
Subject to:
$$\beta \geq 0, \gamma \geq 0, \lambda \geq 0$$

where we set $\log \varphi(\cdot) = const$ [25].

The derivation can be found in Section A of Appendix.

**Remark 7** : The optimization problem (18) can also be derived using hierarchical Bayesian format. Assume $p(w|L) = \mathcal{N}(w|0, L)$, $p(L|P) = (\frac{2}{3})^{p+m}|P|^{\frac{3}{2}}|P - L|^{1/2}$ and $p(P) \propto |P|^{-\frac{5}{2}}$ where $L$ and $P$ are counterparts of $B$ and $\Gamma$ respectively, then $p(w) \propto \int p(w|L)p(L|P)p(P)dLdP$. In SBL, normalized $p(w|y, L, P)$ is used to approximate $p(w|y)$. The value of $L$ and $P$ is evaluated by solving MAP of marginal distribution of hyperparameters as $\max p(L, P|y)$. After simple manipulation, the resultant optimization problem is same with (18). In this hierarchical format, $L$ takes charge of element sparsity of $w$. $p(L|P)$ brings an underlying constraint $0 \preceq L \preceq \Gamma$ which controls sparsity of $L$. $p(P)$ is similar to a Jeffrey prior thus enforcing group sparsity to $L$. Therefore, hyperparameter $P$ imposes group sparsity to $w$ indirectly via $L$.

### 4.3 Algorithm to solve Type II maximization

Since optimization problems $\mathcal{L}_1$ and $\mathcal{L}_2$ are equivalent, we begin with the solution to $\mathcal{L}_1$.

#### 4.3.1 Solve $\mathcal{L}_1$ as a DCP problem

Let

$$\begin{aligned}u(w, \lambda, \beta, \gamma) &= \lambda^{-1}\|y - \Phi w\|_2^2 + w^T(\Gamma^{-1} + B^{-1})w \\ v(\lambda, \beta, \gamma) &= -\log |B + \Gamma| - log|\lambda I + \Phi(\Gamma^{-1} + B^{-1})^{-1}\Phi^T|,\end{aligned} \quad (19)$$

then the optimization problem $\mathcal{L}_1$ becomes:

$$\min_{\gamma \geq 0, \beta \geq 0, \lambda \geq 0} u(w, \lambda, \beta, \gamma) - v(\lambda, \beta, \gamma) \quad (20)$$

**Proposition 8** : Functions $u(w, \lambda, \beta, \gamma)$ and $v(\lambda, \beta, \gamma)$ are both jointly convex with respect to their own variables.

proof: The derivation can be found in Section B of Appendix.

As a result, the cost function actually is a difference of two convex functions and thus is a difference of convex programming (DCP) problem. It can be solved using sequential convex optimization techniques. Here, we use convex-concave procedure (CCCP) which belongs to majorization-minimization (MM) algorithm using the linear majorization function [6,18]. For $\min_x f(x)$ where $f(x) = u(x) - v(x)$ and $u(x), v(x)$ are convex, we can solve it iteratively by:

$$x^{n+1} = \arg \min_x u(x) - <x, \nabla v(x^n)>, \quad (21)$$

where $< \cdot, \cdot >$ denotes inner product.

Therefore, the optimization problem $\mathcal{L}_1$ can be decomposed into sequential convex optimization problems:

$$\begin{aligned}[\gamma^{n+1}&, \beta^{n+1}, \lambda^{n+1}, w^{n+1}] \\ = \arg &\min_{\gamma,\beta,\lambda,w} u(w, \lambda, \beta, \gamma) - \frac{\partial v}{\partial \lambda}\Big|_{[\lambda^n, \beta^n, \gamma^n]}\lambda \\ &- \nabla_\beta^T v\Big|_{[\lambda^n, \beta^n, \gamma^n]}\beta - \nabla_\gamma^T v\Big|_{[\lambda^n, \beta^n, \gamma^n]}\gamma\end{aligned} \quad (22)$$
subject to:
$$\beta \geq 0, \gamma \geq 0, \lambda \geq 0$$

where

$$\begin{aligned}-\frac{\partial v}{\partial \lambda} &= trace\{\Delta\} \\ -\nabla_\beta v &= (\gamma + \beta)^{-1} + diag\{\Phi^T \Delta \Phi\}.^*[\gamma^2.^*(\gamma + \beta)^{-2}] \\ -\frac{\partial v}{\partial \gamma_i} &= \sum_{j=1}^k \frac{1}{\gamma_i + \beta_{ij}} + \frac{\beta_{ij}^2(\Phi^T \Delta \Phi)_{qq}}{(\beta_{ij} + \gamma_i)^2} \\ \Delta &= [\lambda I + \Phi(\Gamma^{-1} + B^{-1})^{-1}\Phi^T]^{-1} \\ q &= (i-1)k + j.\end{aligned} \quad (23)$$

If we optimize $\gamma$, $\beta$ and $\lambda$ first, we get analytical expressions of their optimal solutions as functions of $w$:

$$\beta_{ij}^{opt} = \frac{|w_{ij}|}{\sqrt{g_{ij}^{\beta}}}, \quad \gamma_i^{opt} = \frac{\|w_i\|_2}{\sqrt{g_i^{\gamma}}}, \quad \lambda^{opt} = \frac{\|y - \Phi w\|_2}{\sqrt{g^{\lambda}}}. \quad (24)$$

where

$$\begin{aligned} g_{ij}^{\beta} &= -\nabla_{\beta}^{ij} v|_{[\lambda^n, \beta^n, \gamma^n]} \\ g_i^{\gamma} &= -\nabla_{\gamma}^{i} v|_{[\lambda^n, \beta^n, \gamma^n]} \\ g^{\lambda} &= -\frac{\partial v}{\partial \lambda}|_{[\lambda^n, \beta^n, \gamma^n]}. \end{aligned} \quad (25)$$

Therefore, algorithm (22) can be further simplified by substituting (24):

$$w^{n+1} = \arg\min_{w}$$
$$\sqrt{g^{\lambda}}\|y - \Phi w\|_2 + \sum_{i=1}^{p+m} \sum_{j=1}^{k} \frac{1}{k}\sqrt{g_i^{\gamma}}\|w_i\|_2 + \sqrt{g_{ij}^{\beta}}|w_{ij}|. \quad (26)$$

The optimization problem (26) can be solved as a Second Order Cone Program (SOCP). It has a very similar form with SGL which blends $\ell_1$ and $\ell_2$-norm as the penalty [17]. The differences are an extra weight imposed to the first data-dependent term with 2-norm instead of $\ell_2$-norm thus leading to a very different function property. Such variation is due to the estimation of $\lambda$ incorporated into the problem. If $\lambda$ is known, then the resultant optimization problem is same with standard SGL.

A special attention should be paid to the estimation of the noise variance $\lambda$, which is already mentioned in [20, 24]. Under some circumstance, $\lambda$ can be compensated by two hyperparameters thus unidentifiable. That is mainly caused by limited data points. A simple interpretation is law of large numbers. A large number of data is essential to fully explore the property of a distribution. Therefore, one should not expect to get an accurate estimation of the noise variance but rather treat it as a tuning parameter of the optimization problem. Under this consideration, other reasonable criteria to explore the value of $\lambda$ are recommended such as cross validation.

To summarize, the above procedure can be described as Algorithm 1.

*4.3.2 Solve CCCP by Alternating Direction Method of Multipliers (ADMM)*

If the network is large leading to high dimensional variables, solving (26) will be very slow due to hardware limitation. In this case, we split the optimization problem into a series of independent sub-problems to reduce computation burden. It turns out that the split cost function

---

**Algorithm 1** Solve $\mathcal{L}_1$ using CCCP

1: Initialize $\beta^0$, $\gamma^0$, $\lambda^0$
2: Calculate $g_{ij}^{\beta}$, $g_i^{\gamma}$ and $g^{\lambda}$ using (23) and (25)
3: **for** $n = 1 : Max$ **do**
4:    Solve the reweighted convex problem:

$$w^{n+1} = \arg\min_{w}$$
$$\sqrt{g^{\lambda}}\|y - \Phi w\|_2 + \sum_{i=1}^{p+m} \sum_{j=1}^{k} \frac{1}{k}\sqrt{g_i^{\gamma}}\|w_i\|_2 + \sqrt{g_{ij}^{\beta}}|w_{ij}|$$
(27)

5:    Update $\beta^{n+1}$, $\gamma^{n+1}$ and $\lambda^{n+1}$ according to (24)
6:    Prune out small $\beta_{ij}$ and $\gamma_i$ and remove corresponding columns of dictionary matrix $\Phi$
7:    Update $g_{ij}^{\beta}$, $g_i^{\gamma}$ and $g^{\lambda}$ using (23) and (25)
8:    **if** Any stopping criteria is satisfied **then**
9:       Break
10:   **end if**
11: **end for**

---

can be treated as a sharing problem and can be solved using ADMM algorithms [3, 13].

The optimization problem (26) is firstly transferred into a sharing problem:

$$w_i^{n+1} = \arg\min_{w}$$
$$\sqrt{g^{\lambda}}\|y - \sum_{i=1}^{p+m} z_i\|_2 + \sum_{i=1}^{p+m} \sum_{j=1}^{k} \frac{1}{k}\sqrt{g_i^{\gamma}}\|w_i\|_2 + \sqrt{g_{ij}^{\beta}}|w_{ij}|$$

subject to
$$\phi_i w_i - z_i = 0. \quad (28)$$

where

$$\Phi = \begin{bmatrix} \phi_1 | \ldots | \phi_{p+m} \end{bmatrix}, \\ \phi_i \in \mathcal{R}^{Mm \times k}. \quad (29)$$

Such a problem can be solved using a scaled ADMM method [3]:

$$\begin{aligned} w_i^{n+1} &:= \arg\min_{w_i} \sqrt{g_i^{\gamma}}\|w_i\|_2 + \sum_{j=1}^{k} \sqrt{g_{ij}^{\beta}}|w_{ij}| \\ &\quad + (\rho/2)\|\phi_i w_i - \phi_i w_i^n + \overline{\Phi w}^n - \overline{z}^n + u^n\|_2^2 \\ \overline{z}^{n+1} &:= \arg\min_{\overline{z}} \sqrt{g^{\lambda}}\|y - (p+m)\overline{z}\|_2 \\ &\quad + [(p+m)\rho/2]\|\overline{z} - u^n - \overline{\Phi w}^{n+1}\|_2^2 \\ u^{n+1} &:= u^k + \overline{\Phi w}^{n+1} - \overline{z}^{n+1}, \end{aligned} \quad (30)$$

where
$$\overline{\Phi w}^n = 1/(p+m) \sum_{i=1}^{p+m} \phi_i w_i^n. \quad (31)$$

$w$-update is a standard SGL problem which can be efficiently solved using accelerated generalized gradient descent algorithm [17]. $z$- update is a Group Lasso problem and can be solved analytically. Let $\hat{z} = \bar{z} - \frac{y}{p+m}$, the original $\bar{z}$ update becomes:

$$\hat{z}^{n+1} := \arg\min_{\hat{z}} \frac{1}{2}\|\hat{z} + \frac{y}{p+m} - u^n - \overline{\Phi w}^{n+1}\|_2^2 + \frac{\sqrt{g^\lambda}}{\rho\sqrt{p+m}}\|\hat{z}\|_2. \quad (32)$$

so that

$$\hat{z}^{n+1} = \frac{c}{\|c\|_2}\left(\|c\|_2 - \sigma\right)_+ \quad (33)$$

where

$$\begin{aligned} c &= -\frac{y}{p+m} + u^n + \overline{\Phi w}^{n+1} \\ \sigma &= \frac{\sqrt{g^\lambda}}{\rho\sqrt{p+m}} \end{aligned} \quad (34)$$

Note that $w$-update can be solved in parallel. $\bar{z}$ and $u$-update are then solved in sequence after collecting $w$-update.

### 4.3.3 Solve $\mathcal{L}_2$ using EM method

Now, to solve $\mathcal{L}_2$, first note that it has a similar form with a traditional SBL problem [20, 23] except an extra term $log|B + \Gamma|$ which plays a key role in our formulation. If we remove this term, then by replacing $[(B^{-1} + \Gamma^{-1})^{-1}]$ with a new variable the resultant problem is exactly same with SBL. The most common way to solve SBL are EM methods by treating $w$ as the hidden variable [2,24]. Following the basic procedure of EM method, the algorithm is described in Algorithm 2.

$\mathcal{L}_2$ can also be solved directly by setting the gradient of cost function 0 in order to find stationary points [20]:

$$-\frac{1}{\beta_{ij}^2}(\mu_{ij}^2 + \Sigma_{qq}) + \frac{1}{\beta_{ij}} = 0$$

$$-\frac{1}{\gamma_i^2}\sum_{j=1}^{k}(\mu_{ij}^2 + \Sigma_{qq}) + \frac{k}{\gamma_i} = 0$$

$$\lambda = \frac{\|y - \Phi\mu\|_2^2 + \lambda\sum_{i=1}^{p+m}\sum_{j=1}^{k} 1 - (\beta_{ij}^{-1} + \gamma_i^{-1})\Sigma_{qq}}{N} \quad (37)$$

There is no analytical solution to equations (37). However, we can search for optima by cycling through variables in sequence. Applying simple manipulations to (37) results in the same iteration in Algorithm 2.

**Remark 9** $\mathcal{L}_2$ *can also be solved using CCCP. The first term of the cost function is convex since it is partial minimum of $u(w,\beta,\gamma)$. Clearly, it coincides with Algorithm 1. The only difference comes from the sequence of optimizing variables.*

**Algorithm 2** Solve $\mathcal{L}_2$ using EM
1: Initialize $\beta^0, \gamma^0, \lambda^0$
2: **for** $n = 1 : Max$ **do**
3:     E step: Formulate $p(w|y, \beta^n, \gamma^n, \lambda^n)$ according to (15) and (16)
4:     M step:

$$[\beta^{n+1}, \gamma^{n+1}, \lambda^{n+1}] = \arg\min E_{w|\beta^n,\gamma^n,\lambda^n}\{\ln p(y,w|\beta,\gamma,\lambda)\}. \quad (35)$$

5:     Update solutions of M step as:

$$\begin{aligned} \gamma_i^{n+1} &= \frac{1}{k}\sum_{j=1}^{k}\Sigma_{qq}^n + (\mu_{ij}^n)^2 \\ \beta_{ij}^{n+1} &= \Sigma_{qq}^n + (\mu_{ij}^n)^2 \\ \lambda^{n+1} &= \frac{\|y - \Phi\mu^n\|_2^2 + \lambda^n\sum_{i=1}^{p+m}\sum_{j=1}^{k} 1 - \tau_{ij}\Sigma_{qq}^n}{N}, \end{aligned} \quad (36)$$

where $\tau_{ij} = (\beta_{ij}^n)^{-1} + (\gamma_i^n)^{-1}, q = (i-1)k + j$ and $N = k(p+m)$
6:     Prune out small $\beta_{ij}$ and $\gamma_i$ and remove corresponding columns of dictionary matrix $\Phi$
7:     **if** Any stopping criteria is satisfied **then**
8:         Break;
9:     **end if**
10: **end for**

Although EM method and CCCP both belong to the class of majorization-minimization (MM) method and are special cases of DCA(Difference of Convex functions Algorithm), a DC function has infinite many DC decompositions which can greatly influence the performance of the algorithm [19]. To solve a standard linear regression problem with the dictionary matrix $\Phi \in \mathbb{R}^{N \times M}$ and $N \ll M$, the cost of EM method is $O(MN^2)$. However, if the dimension of $w$ increases either because of the scale of the network or the model complexity, ADMM approach may be more advantageous.

## 5 Simple extension to nonlinear ARX model

Most networks in reality are nonlinear such as genetic regulation networks. To cope with this fact, we introduce nonlinear terms to the multivariable ARX model.

For $i$th node:

$$y_i(t) = f_i(t) + D_{i1}^y(z^{-1})y_1(t) + \ldots + [1 - N_i(z^{-1})]y_i(t) + \ldots + D_{im}^u(z^{-1})u_m(t) + e_i(t). \quad (38)$$

We assume $f_i$ is the linear combination of nonlinear functions which form a dictionary or selected kernel func-tions [14].

$$f_i[y(t), u(t)] = \sum_{j=1}^{q} \alpha_j^i f_{ij}[y(t), u(t)] \quad (39)$$

To select most relevant nonlinear terms, sparse priors are introduced based on the property of those terms. For instance, if a group of nonlinear terms describes the potential transcription activity of a transcriptional factor, then group sparsity can be imposed. In addition, if it is not sure wether a transcriptional factor is repressive or active, then element sparsity is useful to select the right hill functions.

**Remark 10** : *The concept of nonlinear ARX model can be extended to a more general case. In biological system, the ground truth model is always nonlinear. To identify a nonlinear system directly is intractable in most cases and approximating such systems using linear ones is only valid within a small region of dynamics. We hence propose an intermediate model between complete nonlinear and linear system so that it is more informative than linear models but still tractable at the same time. We propose a state space model where all the hidden states (hidden nodes) are described by a linear system while dynamics of observable states (manifest nodes) also includes nonlinear dictionary functions whose variable only involves observable states. Inference of a sparse network described by such a state space model still needs further research. However, one simplified version is nonlinear multivariable ARX model. All the observable states are nodes in ARX model and nonlinear terms are only associated with these nodes.*

## 6 Simulation

We constructed ground truth ARX models with 10 nodes and applied the technique to identify them. To compare with other traditional methods, GSB and SBL were also employed at the same time. The network topology of independent experiment was fixed while polynomial coefficients were generated randomly with maximum order to be 3. We set the upper bound of polynomial order to be 6. The exciting input was designed to be Gaussian random noise. 20 time-series data points were collected for each node and input in each experiment.

We first compare the inference of boolean structure of networks using difference methods in Table 1. The first column records minimum True Positive of multiple experiment, second column maximum False Positive and third column rate of correct inference (100% TP with 0% FP).

To evaluate the estimation accuracy of parameters, we

Table 1
Inference of the network topology of 100 experiment

|  | TP(Min) | FP(Max) | Correct |
|---|---|---|---|
| Our method | 95% | 0% | 99% |
| SBL | 95% | 3% | 83% |
| GSBL | 91% | 0% | 38% |

calculate the infinite norm of estimation error in Table 2:

$$\|x_{err}\|_\infty = \max |x_i| \\ x_{err} = x_{est} - x_{true}. \quad (40)$$

Table 2
Infinite norm of estimation error of 100 experiment

|  | Mean | Min | Max |
|---|---|---|---|
| Our method | 2e-1 | <1e-5 | 1.0 |
| SBL | 4e-1 | <1e-5 | 4.9 |
| GSBL | 1.2 | 9.6e-1 | 2.2 |

Table 1 indicates that our method and GBL successfully exclude redundant correlation among nodes. Our method is superior than the other two by ensuring 99% correct inference of the network topology. GSBL is even worse than SBL in lower rate of TP and total correct inference. Simulation shows that GSBL intends to search for the correct network topology by setting the corresponding groups of parameters 0 but the resultant nonzero groups do not have sparse structure. Hence, the polynomial order is not reduced if a big upper bound is used. Also, GSBL is not able to explore the sparse profile within each group even if the ground truth polynomial coefficient is sparse and the lowest upper bound is used. As a result, its estimation accuracy is lowest among three approaches. In contrast to GSBL, SBL pays more attention to finding accurate estimation of parameters within each group. Its ability to infer the network topology and provide accurate parameter estimation seems satisfactory. However, when applied to real data in practice such as biological time series, one does not expect it to yield group sparse result, which in turn generates full-connected networks. Our method has the best performance since it not only explores the network topology but also pursues the sparsity in each group to find the lowest polynomial order possible. Although sometimes its parameter estimation is biased, the network topology is inferred correctly with high probability. The biased estimation may be due to two possibilities. The major reason is the dictionary matrix correlates with the noise in the model. That means some good properties of SBL when dealing with linear regression problems no long apply in this case [25]. Additional simulation using the same framework to solve standard linear regression problems confirms this concern since the algorithm is capable to recover the weighting vector of a linear regression model. As such, deeper research in the performance of SBL when applied to system identification is required. The second reason is that both CCCP and EM methods only guarantee convergence to a suboptimal solution. Simulation reveals that by changing the initial point, some of the biased estimation can be fixed.

# 7 CONCLUSION AND DISCUSSION

This paper proposes a method to identify the multivariable ARX model. The identification problem is formulated and solved using sparse Bayesian learning. Given measured time series data from the nodes of the network, the method is able to infer the topology and internal dynamics of the network without any prior knowledge, including its topology or system order. The newly proposed method penalises the complexity of the network by introducing both group and element sparsity. These two kinds of sparse priors are combined and its corresponding optimization problem is closely related to Type I Sparse Group Lasso method.

Further development includes two aspects. The first is to find out the theoretical guarantees of the algorithm performance. The second question is how to extend this framework to infer a more general network model describes by a dynamical structural function. The main obstacle here is that SBL normally demands to formulate the problem into a linear regression format so that the likelihood is a quadratic function of parameters. Without this, variational inference or approximate Bayesian computation may be more appropriate.

## A Derivation of Type II maximization problem

Firstly, note that:

$$-2\log \int p(y|w)\hat{p}(w)dw$$
$$-2\log \int \mathcal{N}(y|\Phi w, \lambda I)\mathcal{N}(w|0,B)\mathcal{N}(w|0,\Gamma)\varphi(\beta)\varphi(\gamma)dw$$
$$-2\log \int \exp(E_w)dw + \log|\lambda I| + \log|B| + \log|\Gamma|dw. \quad (A.1)$$

where

$$E_w = -\frac{1}{2}[\lambda^{-1}\|(y-\Phi w)\|_2^2 + w^T(B^{-1}+\Gamma^{-1})w], \quad (A.2)$$

ignoring all the constant terms.

By completing the square:

$$-\frac{1}{2\lambda}\|(y-\Phi w)\|_2^2 - \frac{1}{2}w^T(\Gamma^{-1}+B^{-1})w$$
$$= -\frac{1}{2}\left[(w-\mu)^T\Sigma^{-1}(w-\mu) + E_y\right] \quad (A.3)$$

where

$$\Sigma = \left[(\Gamma^{-1}+B^{-1}) + \lambda^{-1}\Phi^T\Phi\right]^{-1}$$
$$\mu = \lambda^{-1}\left[(\Gamma^{-1}+B^{-1}) + \lambda^{-1}\Phi^T\Phi\right]^{-1}\Phi^T y$$
$$E_y = \min_w \lambda^{-1}\|(y-\Phi w)\|_2^2 + w^T(\Gamma^{-1}+B^{-1})w \quad (A.4)$$
$$= y^T\left[\lambda I + \Phi(\Gamma^{-1}+B^{-1})^{-1}\Phi^T\right]^{-1}y$$

In addition, note that:

$$-2\log \int \exp\left\{-\frac{1}{2}[(w-\mu)^T\Sigma^{-1}(w-\mu)]\right\}dw$$
$$= -\log|\Sigma| \quad (A.5)$$
$$= \log\left|(\Gamma^{-1}+B^{-1}) + \lambda^{-1}\Phi^T\Phi\right|$$

ignoring all the constant terms.

Using (A.3) and (A.5), the integral in (A.1) becomes:

$$-2\log \int p(y|w)\hat{p}(w)dw$$
$$= E_y + \log\left|(\Gamma^{-1}+B^{-1}) + \lambda^{-1}\Phi^T\Phi\right|$$
$$+ \log|\lambda I| + log|B| + \log|\Gamma|$$
$$= E_y + \log\left|I + (\Gamma^{-1}+B^{-1})^{-1}\lambda^{-1}\Phi^T\Phi\right|$$
$$+ \log|(\Gamma+B)| + \log|\lambda I|$$
$$= E_y + \log\left|I + \lambda^{-1}\Phi(\Gamma^{-1}+B^{-1})^{-1}\Phi^T\right|$$
$$+ \log|(\Gamma+B)| + \log|\lambda I|$$
$$= E_y + \log\left|\lambda I + \Phi(\Gamma^{-1}+B^{-1})^{-1}\Phi^T\right| + \log|(\Gamma+B)|$$
$$= \min_w \lambda^{-1}\|(y-\Phi w)\|_2^2 + w^T(\Gamma^{-1}+B^{-1})w$$
$$+ \log\left|\lambda I + \Phi(\Gamma^{-1}+B^{-1})^{-1}\Phi^T\right| + \log|(\Gamma+B)|. \quad (A.6)$$

As a result, we get:

$\mathcal{L}_1:$
$$\min_{\beta,\gamma,\lambda,w} \lambda^{-1}\|(y-\Phi w)\|_2^2 + w^T(\Gamma^{-1}+B^{-1})w$$
$$+ \log|B+\Gamma| + \log\left|\lambda I + \Phi(\Gamma^{-1}+B^{-1})^{-1}\Phi^T\right|$$
Subject to:
$$\beta \geq 0, \gamma \geq 0, \lambda \geq 0.$$
Or: $\quad (A.7)$
$\mathcal{L}_2:$
$$\min_{\beta,\gamma,\lambda} y^T\left[\lambda I + \Phi(\Gamma^{-1}+B^{-1})^{-1}\Phi^T\right]^{-1}y$$
$$+ \log|B+\Gamma| + \log\left|\lambda I + \Phi(\Gamma^{-1}+B^{-1})^{-1}\Phi^T\right|$$
Subject to:
$$\beta \geq 0, \gamma \geq 0, \lambda \geq 0.$$

## B Convexity of DCP problem

To see functions $u(w,\lambda,\beta,\gamma)$ and $v(\lambda,\beta,\gamma)$ are jointly convex functions, we need to prove each term is jointly convex. To check the convexity of $u(w,\lambda,\beta,\gamma)$, we consider the epigraph of its two terms defined as $epif = \{(x,t)|x \in domf, f(x) \leq t\}$ [4]. It is known that:

$$\lambda I \succ 0, \quad \lambda^{-1}\|(y-\Phi w)\|_2^2 < t$$
equivalent to
$$\begin{bmatrix} \lambda I & y-\Phi w \\ (y-\Phi w)^T & t \end{bmatrix} \succ 0 \quad (B.1)$$

so the term $\lambda^{-1}\|(y-\Phi w)\|_2^2$ is jointly convex as is same with $w^T(\Gamma^{-1}+B^{-1})w$ since their epigraph is convex set described by LMI [4].

For the function $v(\lambda,\beta,\gamma)$, firstly note that $-\log|\cdot|$ is a convex function in $S^+$. Since $B+\Gamma$ is an affine function of $\Gamma$ and $B$, $-\log|B+\Gamma|$ is jointly convex with respect to

$\Gamma$ and $B$. The second term of the function $v$ seems more complex. To prove its convexity, we first consider the following lemma. A similar lemma with lower dimension of function domain can be found in [4]:

**Lemma 11** *If a function $f(x) = h(g_1(x), ..., g_k(x))$ where $h(z_1, ..., z_k) : R^k \to R$ and $g_i : R^n \to R$, then $f$ is concave if $h$ is concave and nondecreasing in each argument and $g_i$ are concave.*

The proof of this lemma is straightforward by checking the Hessian matrix:

$$\nabla^2 f = J_g^T \nabla^2 h J_g + \sum_{l=1}^{k} \frac{\partial h}{\partial z_k}\Big|_g \nabla^2 g_l, \quad (B.2)$$

where $J_g$ denotes Jacobian matrix of function $g = [g_1, ..., g_k]^T$.

Let $h(x^1, x) = log|x^1 I + \Phi diag(x)\Phi^T|$ and $g^1(\lambda, \beta, \gamma) = \lambda$, $g_{ij}(\lambda, \beta, \gamma) = \frac{\gamma_i \beta_{ij}}{\gamma_i + \beta_{ij}}$ with $x \in R^{(p+m)k}$, $i \in [1, p+m]$ and $j \in [1, k]$, then $h(g^1, g_{11}, g_{12}, ..., g_{(p+m)k}) = \log|\lambda I + \Phi(\Gamma^{-1} + B^{-1})^{-1}\Phi^T|$. Obviously, $h(\cdot)$ is jointly concave with respect to $x^1$ and $x$.

Since the Hessian matrix of $g^1$ is 0, we only need to check the gradient of $h(\cdot)$ with respect to $x$:

$$\begin{aligned}\frac{\partial h}{\partial x_i} &= trace\left[\Phi^T(x^1 I + \Phi diag(x)\Phi^T)^{-1}\Phi diag(e_i)\right] \\ &= \Delta_{ii},\end{aligned} \quad (B.3)$$

where $\Delta = \Phi^T(x^1 I + \Phi diag(x)\Phi^T)^{-1}\Phi$ and $e_i$ is a vector with its $i$th element 1 and all the others 0.

Since matrix $\Delta$ is at least semi-positive definite, its diagonal elements must be non-negative. As a result, $h(x)$ is nondecreasing in each argument. We finally calculate the Hessian matrix, $H$ of $g_{ij}(\beta, \gamma)$. Note that matrix $H_{qq} = \frac{\partial^2 g_{ij}}{\partial \beta_{ij}^2}$, $H_{ll} = \frac{\partial^2 g_{ij}}{\partial \gamma_i^2}$ and $H_{ql} = H_{lq} = \frac{\partial^2 g_{ij}}{\partial \beta_{ij}\gamma_i}$ with all the other elements 0 where $q = (i-1)k + j + 1$ and $l = (p+m)k + i + 1$. It is always possible to find a permutation matrix such that:

$$P^T H P = \begin{bmatrix} \hat{H} & \mathbf{0} \\ \mathbf{0} & \mathbf{0} \end{bmatrix} \quad (B.4)$$

where

$$\hat{H} = \begin{bmatrix} -\frac{2\gamma_i^2(\gamma_i + \beta_{ij})}{(\gamma_i + \beta_{ij})^4} & \frac{2\gamma_i \beta_{ij}(\gamma_i + \beta_{ij})}{(\gamma_i + \beta_{ij})^4} \\ \star & -\frac{2\beta_{ij}^2(\gamma_i + \beta_{ij})}{(\gamma_i + \beta_{ij})^4} \end{bmatrix}. \quad (B.5)$$

Obviously, matrix $\hat{H}$ is semi-negative definite so that $H$ is also semi-negative definite, which indicates function $g_{ij}(\gamma, \beta)$ is concave. For $g^1$, it is an affine function. According to the lemma above, $log|\lambda I + \Phi(\Gamma^{-1} + B^{-1})^{-1}\Phi^T|$ is jointly concave with respect to $\lambda, \gamma$ and $\beta$.